\documentclass[fleqn,10pt]{SelfArx} 

\usepackage[english]{babel} 
\usepackage[final]{pdfpages}

\definecolor{color1}{RGB}{0,0,90} 
\definecolor{color2}{RGB}{0,20,20} 

\usepackage{hyperref} 
\hypersetup{hidelinks,colorlinks,breaklinks=true,urlcolor=color2,citecolor=color1,linkcolor=color1,bookmarksopen=false,pdftitle={Title},pdfauthor={Author}}

\JournalInfo{~} 
\Archive{} 

\PaperTitle{Neutral theory and scale-free neural dynamics} 

\Authors{\normalsize Matteo Martinello\textsuperscript{1}, Jorge Hidalgo\textsuperscript{1}, Serena di Santo\textsuperscript{2,3}, 
Amos Maritan\textsuperscript{1}, Dietmar Plenz\textsuperscript{4} and Miguel A. Mu\~noz\textsuperscript{2,}*}

\affiliation{\textsuperscript{1}\textit{Dipartimento di Fisica `G. Galilei' and CNISM, INFN, Universit\'a di Padova, Via Marzolo 8, 35131 Padova, Italy}}
\affiliation{\textsuperscript{2}\textit{Departamento de Electromagnetismo y F{\'\i}sica de la Materia e Instituto Carlos I de F{\'\i}sica Te\'orica y Computacional. Universidad de Granada.  E-18071, Granada, Spain.}}
\affiliation{\textsuperscript{3}\textit{Dipartimento di Fisica e Scienza della Terra, Universit\`adi Parma, via G.P. Usberti, 7/A - 43124, Parma, Italy.}}
\affiliation{\textsuperscript{4}\textit{Section on Critical Brain Dynamics, Laboratory of Systems Neuroscience, National Institutes of Mental Health, Bethesda, Maryland, United States of America.}}
\affiliation{*To whom correspondence should be addressed: mamunoz@onsager.ugr.es.} 

\Keywords{Neural avalanches --- Criticality --- Neutral theory -- Generic scale invariance} 
\newcommand{\keywordname}{Keywords} 

\Abstract{Avalanches of electrochemical activity in brain networks have been
  empirically reported to obey scale-invariant behavior
  --characterized by power-law distributions up to some upper
  cut-off-- both {\emph{in vitro}} and {\emph{in vivo}}. Elucidating
  whether such scaling laws stem from the underlying neural dynamics
  operating at the edge of a phase transition is a fascinating
  possibility, as systems poised at criticality have been argued to
  exhibit a number of important functional advantages.  Here we employ
  a well-known model for neural dynamics with synaptic plasticity, to
  elucidate an alternative scenario in which neuronal avalanches can
  coexist, overlapping in time, but still remaining
  scale-free. Remarkably their scale-invariance does not stem from
  underlying criticality nor self-organization at the edge of a
  continuous phase transition.  Instead, it emerges from the fact that
  perturbations to the system exhibit a neutral drift --guided by
  demographic fluctuations-- with respect to endogenous spontaneous
  activity. Such a neutral dynamics --similar to the one in neutral
  theories of population genetics-- implies marginal propagation of
  activity, characterized by power-law distributed causal
  avalanches. Importantly, our results underline the importance of
  considering causal information --on which neuron triggers the firing
  of which-- to properly estimate the statistics of avalanches of
  neural activity.  We discuss the implications of these findings both
  in modeling and to elucidate experimental observations, as well as
  its possible consequences for actual neural dynamics and information
  processing in actual neural networks.}

\begin{document}
\flushbottom 

\maketitle 

\thispagestyle{empty} 

\section*{Introduction}
The introduction by Kimura in 1968 of the neutral theory
--hypothesizing that most evolutionary change is the result of genetic
drift acting on neutral alleles \cite{Kimura}-- caused much debate and
a revolution in the way population genetics and molecular evolution
were understood. In a similar endeavor, Hubbell proposed that most of
the variability in some ecological communities could be ascribed to
neutral dynamics of similar species which expand or decline as a
result of stochasticity \cite{Hubbell,Review-neutral}.  Neutral
theories have in common that they neglect any \emph{a priori}
intrinsic difference between coexisting individuals, regardless
  of their ``species'' (allele, tree,...) type, implying that the
  dynamics is purely driven by random demographic effects. For
  instance, the introduction of a novel species within an
established population triggers a random cascade of changes, or
``avalanche'', which --as a result of the implicit neutrality-- does
not have an inherent tendency to neither shrink nor to expand at the
expenses of others. This marginal-propagation process generates
scale-free avalanches, which resembles critical ones even if the
system is not necessarily posed at the edge of a phase transition
\cite{Review-neutral,Pinto} (a brief mathematical summary of neutral
theory can be found in SI Appendix S1).  Neutral models have been
successfully employed to explain the emergence of scale-free
distributions in (i) epidemic outbreaks with neutral microbial strains
\cite{Pinto}, (ii) viral-like propagation of neutral memes
\cite{Gleeson}, (iii) the evolution of the microbiome \cite{Biome},
and (iv) the renewal of the intestinal epithelium from neutral stem
cells \cite{Simons}.  Could neutral theory be applied to neural
dynamics of the brain?  And, in particular, could it explain the
emergence of neuronal avalanches reported for spontaneous activity?

The human brain has a special feature that is common to all
mammalians: it is endogenously active; i.e. cascades of
electrochemical activity at multiple timescales spontaneously
pervade its dynamical state even in the absence of any apparent
stimuli or task.  Mounting evidence suggests that such an endogenous
activity is not random, but structured, and it contributes
significantly to stimulus-related responses, being essential to brain
functioning. Specifically, spontaneous, spatiotemporal bursts of
neural activity were reported to appear in the form of avalanches
\cite{BP2003}, whose sizes $s$ and durations $t$ are distributed as
$P_s(s) \sim s^{-\tau} ~~ {\cal{F}}(s/s_c)$, and $P_t(t) \sim
t^{-\alpha} ~~ {\cal{G}}(t/t_c)$, respectively, where $\tau \approx
3/2$ and $\alpha\approx 2$ are critical exponents similar to those
of an unbiased branching process \cite{Harris,Marro,SOBP}, $\cal{F}$
and $\cal{G}$ are scaling functions and $s_c$ and $t_c$ are
system-size dependent cut-offs obeying finite-size scaling
\cite{Binney}.  
Similar results have been obtained both {\emph{in vitro}} and
{\emph{in vivo}}, as well as for different tissues, preparation
types, experimental techniques, and animal species (see e.g.
\cite{Peterman2009,Torre2007,Haimovici,Taglia,Plenz-Shriki2013,2-photon,Linken2012}).
Remarkably, signs of scale-invariance have been reported to vanish
under abnormal circumstances such as under the influence of modified
pharmacological conditions, under anesthesia, or in pathological
conditions \cite{Healthy}. We refer to
\cite{Schuster-book,Chialvo2010,Plenz-2012,Beggs-criticality,Lucilla2014}
for overviews and discussions on the state of the art.  Taken
together, these observations suggest that scale-free avalanches are
a generic feature of spontaneous activity in cortical tissues,
suggesting that they stem from an underlying critical phenomenon
(see however \cite{Touboul1}), and this conclusion seems to back the
hypothesis that biological computing systems might operate at the
edge of phase transitions \cite{Langton1990,Ber-Nat,Hidalgo2014},
providing them with optimal transmission and storage of information,
exquisite sensitivity to signals, and a number of other important
functional advantages \cite{Plenz-Functional,Kinouchi-Copelli}.

Scale-free distributed events or bursts of ``activity'' such as
earthquakes, vortex avalanches in superconductors, and Barkhaussen
noise are common place in Nature (see e.g. \cite{Bak,Sethna2001})
and are often ascribed to their underlying dynamics being poised at
a critical point. The paradigm of ``self-organized criticality'' was
developed to explain how and why natural systems could self-tune to
the vicinity of critical points \cite{Bak,Jensen,Pruessner-book};
scale-free distributed avalanches turn out to be the fingerprint of
critical points of a phase transitions into quiescent (or
``absorbing'') states \cite{BJP,JABO1}.  Despite the success and
conceptual beauty of this framework, not all scale-invariant
episodes of activity can be ascribed to underlying criticality
\cite{GG,Newman-powerlaws}; for instance, power-law distributed
excursion sizes and times are generated by unbiased random walks
\cite{Redner}, self-organization to the edge of a discontinuous
phase transition \cite{Serena-PRL} and, as discussed above, neutral
dynamics \cite{Pinto,Simons}.

In this paper we explore the possibility that empirically observed
neural avalanches could be scale-free as a result of an underlying
neutral dynamics --i.e. that each single event of activity is
indistinguishable from others and can marginally propagate through
the network-- alternatively to being self-organized to the edge of a
phase transition. This is, we explore whether scale-free
avalanches could stem from the neutral competition for available
space of activity generated from different sources or stimuli.  
We put forward a subtle but important difference between 
such causal avalanches and existing empirical evidence, 
and discuss how neutral patterns of activity --i.e. coexisting
causal avalanches of many different shapes, sizes and durations--
could be exploited by real neural systems for efficient coding,
optimal transmission of information and, thus, for memory
and learning \cite{abbott2000},

\section*{Results}
\subsection*{Computational model and its phenomenology}

\begin{figure}[tb!]
\begin{center}
\includegraphics[width=\columnwidth]{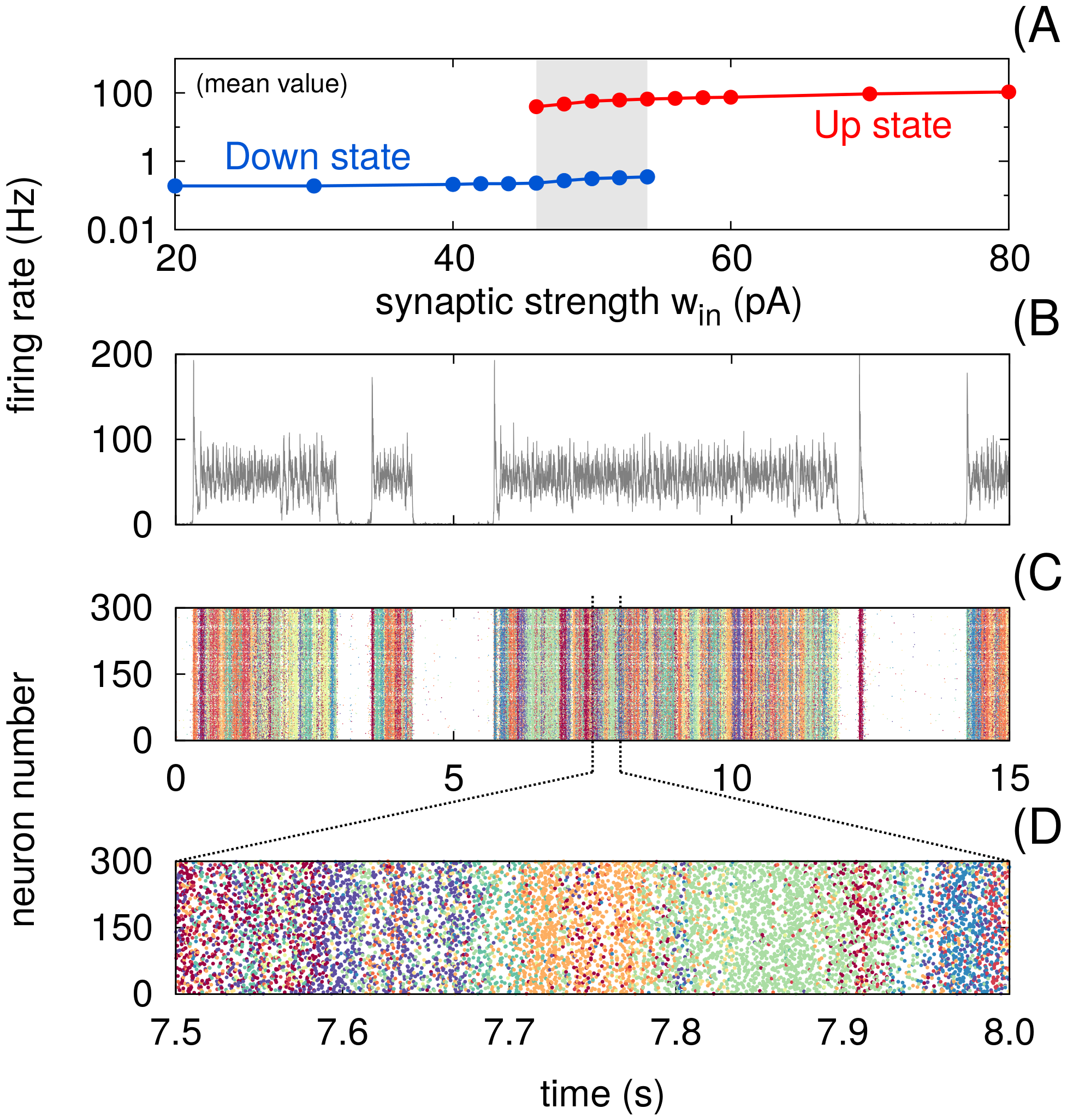}
\caption{ (Color online) Numerical integration of the model of Millman
  \emph{et al.} \cite{Millman} with $N=300$ neurons. (A) Phase
  diagram of the mean firing rate as a function of the synaptic
  strength parameter, $w_\mathrm{in}$.  For low values of
  $w_\mathrm{in}$, the stable state is a quiescent state with very low
  levels of activity (Down state), whereas for large values of
  $w_\mathrm{in}$, the system exhibits high levels of activity (Up
  state). Both states coexist for intermediate strength values (shaded
  region), allowing for Up-and-Down transitions. Importantly, the
  transition is discontinuous. (B) Timeseries of the network
  firing rate for $w_\mathrm{in}=50$ pA illustrate the system's
  bistability, with eventual (stochastic) jumps between Up and Down
  states. (C) Raster plot (for the same times as above)
  in which distinct colors are used for different causal  avalanches,
  defined as subsequently activated neurons after a spontaneous
  activation of a neuron by an
  external input \cite{Millman}. (D)Raster plot zoom (broken
  lines) demonstrating the intermingled and temporally
  overlapping organization of different causal avalanches.  Model
  parameters have been set as in \cite{Millman} (see MM).
  \label{fig:patterns}}
\end{center}  \end{figure}

Using a model of leaky integrate-and-fire neurons regulated by
synaptic plasticity, Millman, \emph{et al.}  \cite{Millman} were able
to capture the empirical observation of bistability in cortical
networks, i.e. two well differentiated stable patterns of cortical
activity, called Up and Down states (see e.g.
\cite{Steriade,Hidalgo2012} and Refs. therein). Briefly, the model
consists of $N$ leaky integrate-and-fire excitatory neurons forming a
directed random Erd{\H{o}}s-R{\'e}nyi network with average
connectivity $K$. Neurons integrate synaptic inputs from other neurons
and fire action potentials, which rapidly deplete associated synaptic
resources. These latest recover at a slow time scale, thereby
regulating the overall level of activity in the network (see MM).  The
model can be tuned by controlling e.g. its average synaptic
strength. For weak synaptic strengths, a quiescent phase with very low
levels of activity, the Down state, exists, whereas a second, stable
state with high firing rates, the Up state, emerges for large synaptic
strengths (see Fig. \ref{fig:patterns}A). For intermediate
strengths, spontaneous fluctuations allow for rapid Up and Down states
alternations (see Fig.  \ref{fig:patterns}B). This
phenomenology -- which can also be reproduced by keeping synaptic
strength fixed and varying the synaptic recovery time or some other
parameter of the model-- corresponds to a discontinuous phase
transition (see Fig. \ref{fig:patterns} and Fig. 1 in \cite{Millman})
and therefore lacks the critical point characteristic of continuous
transitions.  Remarkably, when tracking cascades of neuronal firing
based on participating neurons, i.e. causal avalanches (see below),
the model was shown to exhibit scale-free distributions of sizes and
durations during Up-states, with associated exponents $\tau \approx
3/2$ and $\alpha \approx 2$, i.e. the hallmark of actual neuronal
avalanches. Accordingly, the authors considered the Up state as
``self-organized critical'', in contrast to the Down state which was
``subcritical'' with causal cascades that were not scale-free
\cite{Millman}.  Given that critical dynamics emerge at continuous
phase-transitions, the presence of scale-invariant Up state avalanches
in the absence of any such transition in this model is puzzling, and
prompted us to identify possible alternative mechanisms for the
emergence of scale-free avalanches.

\subsection*{``Causal'' avalanches}
Following \cite{Millman}, we tracked causal
cascades/avalanches.\footnote{Here, we use indistinctly the terms
``avalanche'' and ``cascade''.} Each one is initiated when an
external input depolarizes a neuron's membrane potential above its
threshold to fire an action potential, it unfolds as the membrane
potential of a subsequent neuron surpasses its threshold as a result
of a synaptic input from an existing cascade member, and stops when
this does not happen.  The size of a cascade is the total number of
action potentials it triggered, while the cascade duration is the
timespan between its initiation and the time of its last action
potential \cite{Millman}.  Avalanches were analyzed separately for Up
and Down states in a network with $N=3000$ neurons, using different
values of the external firing rate, $f_e$; in particular we analyzed
the slow driving limit $f_e\rightarrow0$.  Our results are in perfect
agreement with the phenomenology found in \cite{Millman}: cascades in
the Down state do not exhibit scale invariance but instead have a
characteristic scale (see SI Appendix S2). In contrast, cascades
during Up-states distribute in size and duration according to
power-laws with exponent close to $\tau = 3/2$ and $\alpha = 2$,
respectively (see Fig. \ref{fig:millman}A). As already observed in
\cite{Millman}, these results are quite robust, do not depend on how
deep into the Up state --i.e. how far from the transition point--
simulations are run, nor do they change upon introducing inhibitory
neurons (see SI Appendix S3).

\subsection*{``Time-correlated'' avalanches from time binning}
\begin{figure}[tb!]
\begin{center}
\includegraphics[width=\columnwidth]{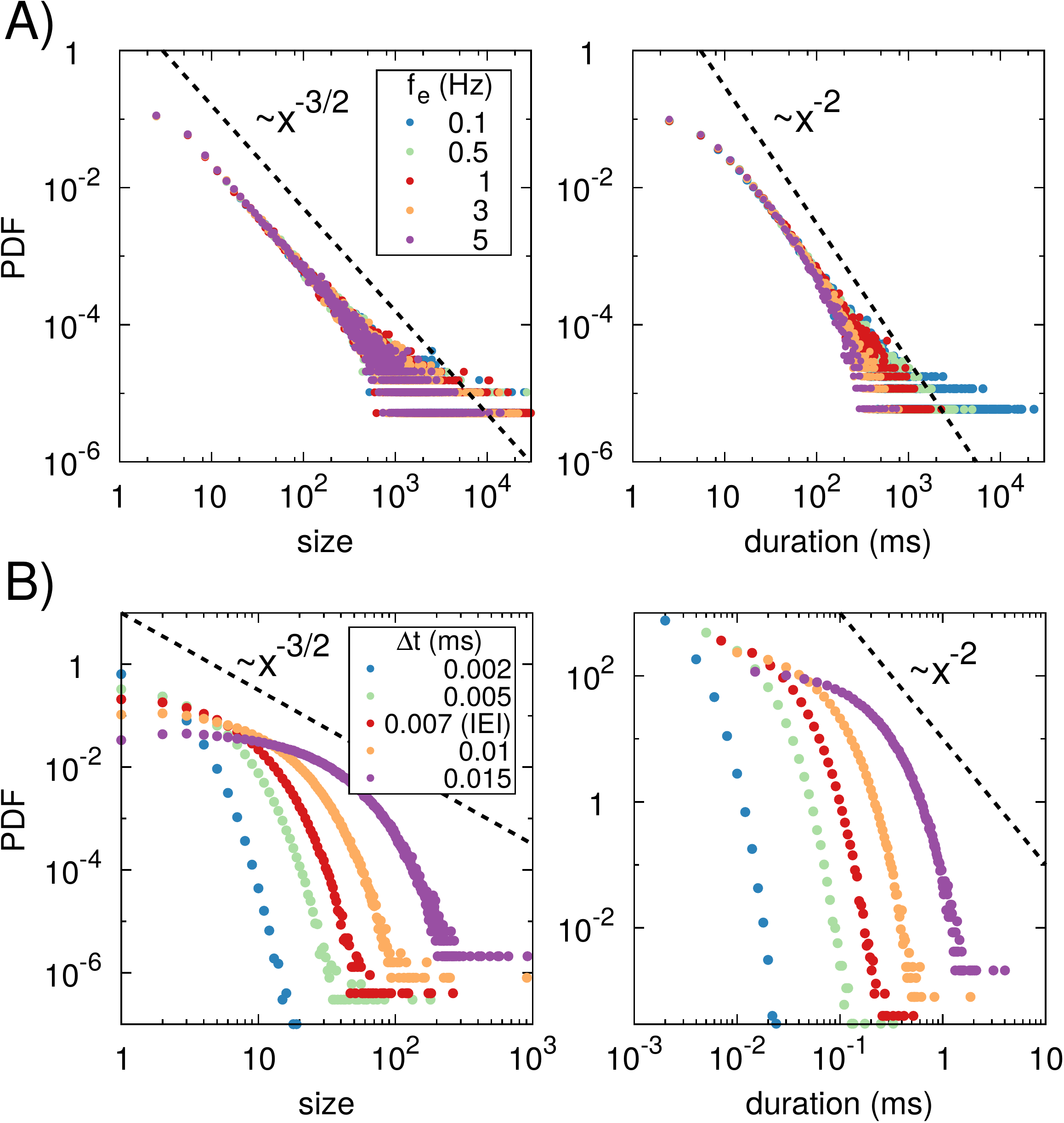}
\caption{(Color online) Avalanche size and duration distributions 
within the Up-state phase in the model of Millman \emph{et al.} \cite{Millman} using two
different methods (double logarithmic plot). (A)
Causal avalanches were defined using the same criterion
as in \cite{Millman}, for several values of the external input
$f_e$, confirming the observation that sizes and durations are
power-law distributed with the same exponents of an unbiased branching
process, i.e. $\tau=3/2$ and $\alpha=2$, respectively
\cite{Harris,SOBP}. (B)``Time-correlated'' 
avalanches, defined with the standard temporal binning method
\cite{BP2003} (which estimates causality by temporal proximity),
using five different time intervals $\Delta t$ to bin the data,
including one coinciding with the average interevent interval (IEI)
as usually done in the analyses of empirical data \cite{BP2003}, for
$f_e=5$ Hz; in this case distributions do not obey a power-law
distribution but have a characteristic scale.  In all cases,
simulations were performed in a network of $N=3000$ neurons (model
parameters as in \cite{Millman}, see MM section).
\label{fig:millman}}
\end{center}
\end{figure}

A key point of the previous analysis is that causal information
between activation events (i.e. ``who triggers who'') is essential
to define avalanches.  However, in empirical analyses it is not
clear whether events occurring nearby in time --usually ascribed to
the same avalanche in statistical analyses-- are actually causally
connected or not.  The standard approach, that has been successfully
used in the analysis of experimental data, where causal information
of event propagation is typically not accessible \cite{BP2003,
  Peterman2009}, consists in defining cascades from a series of
discrete supra-threshold events, by choosing a discrete time bin
$\Delta t$. An avalanche is defined as a sequence of successive time
windows with at least one event in each that is preceded and ended
by an empty bin. In principle, one could expect different scaling
relations when varying the time window $\Delta t$, as was
demonstrated in analyses of empirical data \cite{BP2003}. For
comparison, and following \cite{BP2003,Peterman2009}, we take
$\Delta t$ to be equal to the average inter-event interval (IEI),
defined as the average time interval between successive
events.\footnote{Even though the IEIs can vary for different
  experimental situations, size and duration distributions have been
  claimed to exhibit universal behavior with exponent $\tau \approx
  3/2$ and $\alpha \approx 2$, respectively, provided that data are
  binned using the IEI.}  Using this binning procedure in timeseries
from the computational model, we find that cascade duration and size
distributions obtained from Up states are exponentially distributed
with a characteristic scale, showing no signs of scale-invariant
behavior (see Fig. \ref{fig:millman}B). Distributions did not change
qualitatively for different values of $\Delta t$. Thus, in the model
of Millman \emph{et al.}, cascades based on temporal proximity
differ significantly from cascades based on causal information.
This finding is in contrast to the established scale-free avalanche
distributions that emerge from experimental data based on temporal
proximity.

In the model of Millman \emph{et al.}, causal avalanches can (and
do) coexist in time (see Fig. \ref{fig:patterns}C and D); thus, the
temporal proximity approach does necessarily fail to uncover true
causal avalanches. Thus, our observations, together with the lack of
a continuous phase transition, question the origin of true scale
invariance within Up states and its actual relationship with
experimental findings. To shed light on this problem, in the next
section we analyze a minimal model which captures the main
ingredients for activity propagation, showing that the observed
scale-free causal avalanches in the model of Millman \emph{et al.}
stems from an underlying neutral dynamics \cite{Pinto,Simons}.

\subsection*{Neutral (causal) avalanches in a minimal model for
  activity propagation}
In archetypical models of activity propagation such as the contact
process, directed percolation and the
susceptible-infected-susceptible model \cite{Marro,Henkel},
``active'' sites propagate activity to their nearest neighbors or
become de-activated at some rates. As a result, depending on rate
values, there exist a quiescent and an active phase, as well as a
critical point separating them \cite{Marro,Henkel}; avalanches
started from a single initial event exhibit scale invariance only at
criticality (see SI Appendix S4), and if they are triggered at a
sufficiently slow rate, they do not overlap.

In contrast, within the framework of neutral dynamics, multiple
avalanches can propagate simultaneously. The difference between
critical and neutral avalanches can be vividly illustrated by
considering a variant of the contact process, consisting of many
different but equivalent ``species''. It can be studied arbitrarily
far from the phase transition to explore the statistics of causal
avalanches.  More specifically, we consider a fully-connected
network with $N$ nodes that can be either active ($A$) or inactive
($I$). At every time, each single active site is assigned to a
unique individual avalanche/species $k$ (the one from which it
derives) and labeled by $A_k$. The dynamics is as follows: i) a new
avalanche, with a new label, is initiated by the spontaneous
activation of an inactive site at small driving rate $\epsilon$; ii)
active sites propagate the activity to neighboring inactive places
at rate $\lambda$, and iii) active sites become inactive at rate
$\mu$. This is equivalent to the following set of reactions for
$k=1,...,M(t)$:
\begin{equation}
\begin{array}{ccc}
I &\overset{\epsilon}{\longrightarrow}& A_{M(t)+1}\\
A_k + I &\overset{\lambda}{\longrightarrow}& A_k + A_k\\
A_k &\overset{\mu}{\longrightarrow}& I
\end{array}
\label{eq:reaction}
\end{equation}
where $M(t)$ is the total number of avalanches triggered up
to time $t$.  This dynamical process is \textit{neutral} (or
symmetrical) among avalanches/species as parameter rates do not depend
on label $k$ (see SI Appendix S1 for an extended discussion on neutral
theories).  The duration (resp. size) of an avalanche $k$ is the time
elapsed (resp. total number of activations) between its spontaneous
generation and the extinction of its label. Observe that different
avalanches can coexist (all the most in the active phase) and that the
total number of coexisting avalanches can vary in time.  The state of
the system is determined by $M(t)$ and the number of
$k-$type active sites, $n_k(t)$, or, equivalently, the corresponding
density $\rho_k(t) = n_k(t)/N$.  The total density of active sites is
defined as $\rho(t)=\sum_{k=1}^{M(t)} \rho_k(t)$.
Importantly, the system of Eq. \eqref{eq:reaction} is nothing but the
standard contact process (with a non-vanishing rate for spontaneous
activation $\epsilon$) if avalanche labels are ignored.  Therefore, in
the limit $\epsilon\rightarrow0$, it exhibits a continuous phase
transition for the total activity density at the critical point given
by $\lambda_c=\mu$ \cite{Marro,Henkel}.

We performed computer simulations of the dynamics described by
Eq. \eqref{eq:reaction} by means of the Gillespie algorithm
\cite{Gillespie} in a fully-connected network of size $N=10^4$.
Parameter values are chosen for the system to lie well inside the
active phase, $\lambda=2$, $\mu=1$ (i.e. $\lambda=2\lambda_c$), and
$\epsilon$ taking small values such as $10^{-1}, 10^{-2}, 10^{-3}$ and
$10^{-4}$.  Typical timeseries for individual avalanches, $\rho_k$, as
well as for the total activity, $\rho$, are depicted in
Fig. \ref{fig:CP}A.  Observe that the steady-state overall density
(gray color) coincides, on average, with that of the contact process
in the infinite size limit, $\rho^* \simeq
(1-\mu/\lambda)+\epsilon\mu/(\lambda(\lambda-\mu))$ (see MM).  On the
other hand, individual avalanches (colored curves in
Fig. \ref{fig:CP}A) experience wild fluctuations as a function of
time. The statistics of avalanches is illustrated in
Fig. \ref{fig:CP}B revealing that avalanche sizes and durations are
power-law distributed with exponents $\tau=3/2$ and $\alpha=2$ in the
limit of small spontaneous activation rate $\epsilon\rightarrow0$.
Remarkably, scale-free avalanches appear all across the active phase,
$\lambda>\lambda_c$ (see SI Appendix S4).

\subsection*{Analytical approach}
\begin{figure}[tb!]
\begin{center}
\includegraphics[width=\columnwidth]{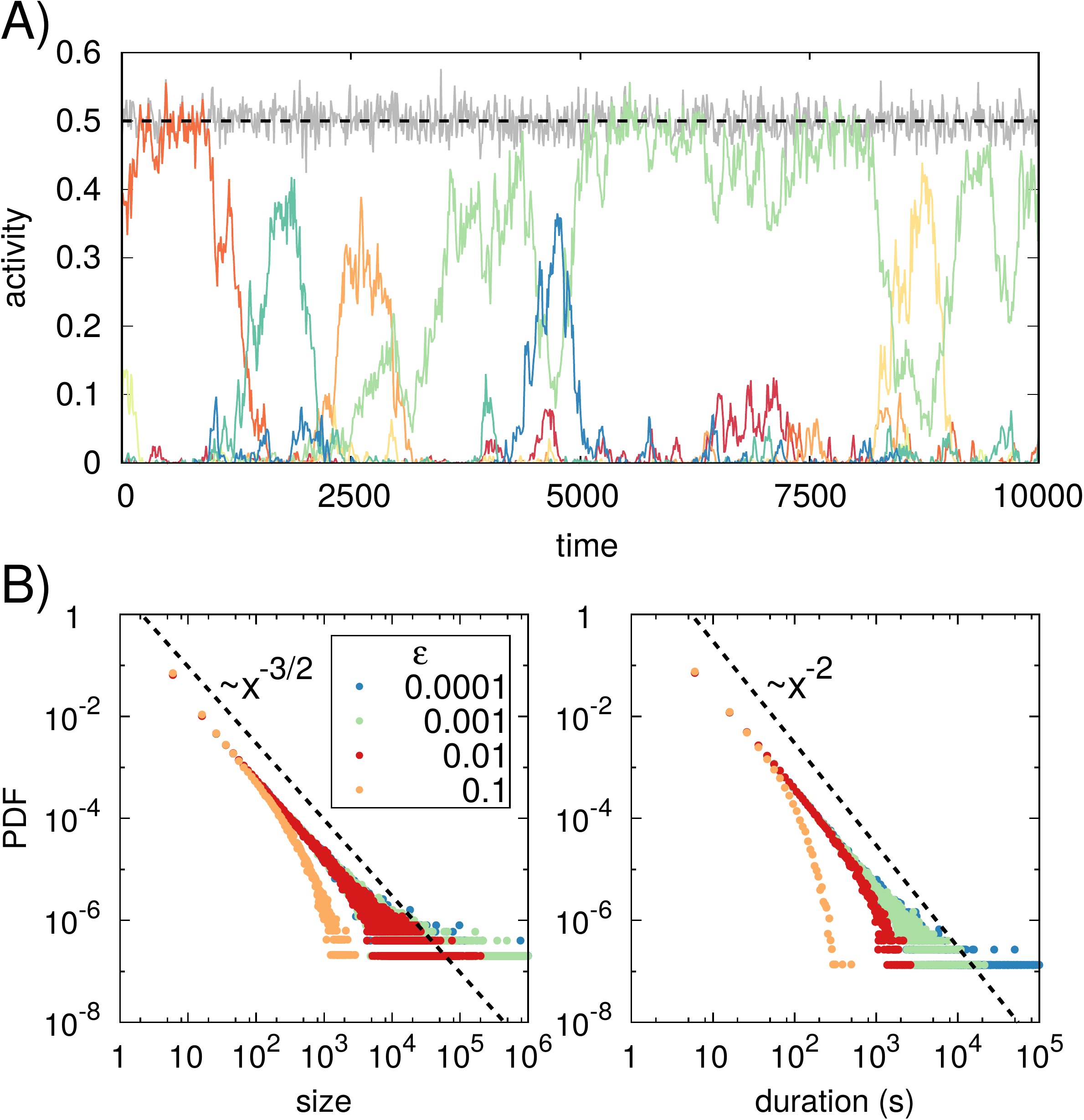}
\caption{ (Color online) Causal avalanches in a minimal model for
  propagation activity, defined as cascades of events initiated from
  the spontaneous activation of one unit, without overlap between
  avalanches (i.e. a given node cannot be simultaneously part of
    more than one avalanche).  (A)  The activity of each
  avalanche is defined as the density of active elements in the system
  belonging to that avalanche, identified with different colors in the
  plot.  The global activity density is represented with the
  gray-colored line.  Parameters of the model are taken deep inside
  the active phase, $\lambda=2$, $\mu=1$, for a system size $N=10^4$
  and small spontaneous activation rate $\epsilon=10^{-3}$. Whereas
  the global activity exhibits slight fluctuations around its
  steady-state value $\rho^*\simeq1-\mu/\lambda$ (represented by the
  dashed line), individual avalanches can exhibit wild variability.
  (B) Avalanche size and duration distributions for different
  values of $\epsilon$ (other parameters as in (A), i.e. deep
  inside the active phase).  Avalanche statistics exhibit robust
  power-law scaling with the same exponents of the neutral theory for
  avalanche propagation (marked with dashed lines for comparison).
\label{fig:CP}}
\end{center}
\end{figure}

To shed light on this result, we study analytically this simplified
model in the large network-size limit. Starting from the master
equation associated to Eq. \eqref{eq:reaction}, performing a
system-size expansion for large but finite system sizes
\cite{Gardiner}, the dynamics of a just-created avalanche is
well-described by the following equation:
\begin{equation}
\dot \rho_k = \left(\lambda(1-\rho)-\mu\right)\rho_k 
+\sqrt{\frac{1}{N}\left(\lambda(1-\rho)+\mu\right)\rho_k} \,\xi_k(t), 
\label{eq:langevin-rhok}
\end{equation}
with the initial condition $\rho_k=1/N$, and where $\xi_k(t)$
represents a zero-mean Gaussian white noise of unit variance (to be
interpreted in the It\^{o} sense \cite{Gardiner}).  If the system is
very large, and \textit{when the spreading rate lies within the
  active phase} ($\lambda>\mu$), the total activity density exhibits
very small fluctuations, remaining quite stable around the
steady-state value, as illustrated by the gray-colored timeseries in
Fig. \ref{fig:CP}A.

To understand this variability, let us assume $\rho(t)\simeq\rho^*$
in Eq. \eqref{eq:langevin-rhok} and $\rho_k\ll\rho$ (which is good
approximation for large system sizes where many avalanches coexist);
thus,
\begin{equation}
\dot \rho_k = -\frac{\mu}{\lambda-\mu}\epsilon \rho_k+
\sqrt{\frac{\mu}{N}\left(2-\frac{\epsilon}{\lambda-\mu}\right) \rho_k} \,\xi_k(t).
\label{eq:langevin-rhok2-i}
\end{equation}
In the limit of small driving, $\epsilon\rightarrow0$, the
deterministic/drift term in Eq. \eqref{eq:langevin-rhok2-i} vanishes,
and the dynamics of avalanche $k$ can be simply written as:
\begin{equation}
\dot \rho_k = \sqrt{\rho_k} ~\xi_k(\hat{t}),
\label{eq:langevin-rhok2}
\end{equation}
where for simplicity in the notation, a factor $2\mu/N$ has been
reabsorbed into the time scale $\hat{t}$.
Eq. \eqref{eq:langevin-rhok2} represents a freely-moving random-walk
with demographic fluctuations, and --as further discussed in SI
Appendix S1-- it describes the evolution of a species density in any
neutral-type of dynamics. In other words, once an avalanche starts,
its statistics is entirely driven by neutral demographic fluctuations,
\textit{regardless of the distance to the critical point}\footnote{Not
surprisingly, Eq. \eqref{eq:langevin-rhok2} corresponds also to the
mean-field description of an unbiased branching process
\cite{Harris}.}. Furthermore, the avalanche exponents associated
with this neutral, noise-driven, dynamics are $\alpha=2$ and
$\tau=3/2$.  Actually, the previous reasoning holds all across the
active (Up) phase; on the other hand, in the quiescent (Down) state,
the steady state activity $\rho^*$ goes to $0$, as the deterministic
driving force in Eq. \eqref{eq:langevin-rhok} is negative, leading to
subcritical avalanches, as indeed reported in \cite{Millman}.

Thus, a simple approach allowed us to explicitly show that neutral
dynamics among coexisting dynamically-indistinguishable avalanches
leads to scale-free distributions all across the active (Up) phase,
i.e. arbitrarily far away from the edge of the phase transition. The
same conclusion extends to the model of Millman \emph{et al.} even if
detailed analytical calculations for such a case are more difficult to
perform.

\section*{Discussion}

A remarkable observation --that has elicited a great deal of
interest-- is that neural activity in the brain of mammals, including
humans, occurs in the form of neuronal avalanches consisting of
outbursts of neural activity intervened by periods of relative
quiescence, across many resolution scales in a robust way
\cite{BP2003,Schuster-book}. For {\emph{in vitro}} studies of
relatively small networks it seems plausible to assume that events
occurring during one of such outbursts are causally connected, so that
activity emerges at some location and transiently propagates through
the network, causing a cascade of co-activations.  However, there is
no clear empirical validation that this is actually the case; diverse
causally-connected cascades could, in principle, occur simultaneously,
hindering their experimental discrimination as individual avalanches.
Obviously, the situation is much more involved in large neural
networks as analyzed {\emph{in vivo}} as diverse scales of resolution,
e.g. from local field potential measurements, magneto-encephalography,
functional magnetic resonance imaging, etc.  There is no known
empirical procedure to actually disentangle causal influences, nor to
discern whether different causal cascades of activations overlap (as
they probably do in functional brains). Thus, in the absence of a
better indicator, events of activity are customarily clustered
together as individual avalanches, relying on a criterion of temporal
proximity.

It thus remains to be fully elucidated what is the true nature of
scale-free avalanches in actual neural systems. To shed light on this,
here we scrutinized the most commonly referred model --introduced by
Millman and coauthors \cite{Millman}-- justifying the emergence of
power-law distributed avalanches in networks of integrate-and-fire
neurons with synaptic plasticity.  First of all, we reproduced the
findings in \cite{Millman}, and confirmed that the model exhibits two
different phases in parameter space, an Up-state characterized by
large average firing rates and a Down-one with small firing, separated
by a discontinuous phase transition.  We carefully analyzed the
dynamics within the active phase, and corroborated that diverse
avalanches can coexist, and that their sizes and durations are
scale-free (with exponents, $3/2$ and $2$, respectively) if and only
if precise information on which neuron triggers the firing of which
--which is accessible in computational models-- is used to identify
(causal) avalanches.  On the other hand, a different analysis --which
is the one customarily applied to empirical data-- based on defining
avalanches through a time-binning procedure, blind to detailed causal
information between activation events, does not reveal any trace of
scale-freedom in avalanche distributions.

These observations naturally pose two important questions. First, if
this model is not self-organized to the edge of a phase transition,
where do the computationally-reported scale-free (causal) avalanches
within this model stem from?  And second, does this model constitute a
faithful representation of actual neural dynamics, including the
experimentally observed scale-invariant avalanches?

To answer the first question we designed a simplified dynamical model
with an overall phenomenology very similar to that of the model of
\cite{Millman}: i.e.  it exhibits scale-invariant causal avalanches
all along its active phase, regardless of the distance to a
phase-transition point (which actually can be either a continuous or a
discontinuous one depending on model details). This simplified model
--a variant of the contact process with many different types of active
particles--- allowed us to uncover that scale-invariant avalanches
within the active phase stem from the neutral dynamics among diverse
coexisting (causal) avalanches.  In particular, if new seeds of
activity are injected at a very slow rate in a system with recurrent
background activity (i.e. in its active phase) each one does not have
a net drift toward contracting or expanding in the background of
recurrent activity in which it unfolds; its dynamics just follows
demographic fluctuations, much as in neutral theories of population
genetics. Moreover, the branching ratio is equal to unity, and causal
avalanches are power-law distributed (as in the unbiased branching
processes) without the model being posed at the edge of a phase
transition.  In summary, the observed scale-invariance in a
well-accepted computational model for neuronal dynamics as well as in
a simplified model stems from the neutrality or symmetry between
diverse competing cascades of causally-related events which coexists
in a background of recurrent activity.

In what respects the second question above, it might occur that the
discussed computational model does not reproduce all the phenomenology
of actual neural dynamics in real networks. For instance, activity
exhibits clear temporal clustering (so that measured power-laws
disappear when times are reshuffled \cite{BP2003}) and, as we have
shown, this fact is mostly lacking in the model of Millman \emph{et
  al.}.  This drawback was overcome in a more recent and detailed
computational model including many additional neuro-physiologically
realistic ingredients (such as e.g. inhibitory plasticity) which
exhibits temporal clustering of activity together with scale-free
avalanches \cite{Plenz2015}.  In this case, avalanches are also
measured employing causal information so that scale-invariance is
likely to stem from underlying neutrality, rather than criticality. It
would be highly desirable to have a study of purely time-binned
avalanches in this type of approach, allowing to compare them with
causal ones.  From a broader perspective, more complete computational
models and/or analyses allowing to scrutinize the possible emergence
and interplay between neutrality and criticality are highly needed.

Finally, the main question that remains to be answered is: given that
various types of functional advantages are ascribed to criticality, do
these same advantages still exist if neuronal scale-free avalanches
turn out to be the consequence of underlying neutrality rather than of
the tuning to the edge of a phase transition?  While we do not have a
definite answer to this, we speculate that this type of power-law
distributed coexisting causal avalanches could play a fundamental role
in neural functioning. In particular, there are known biological
mechanisms, such as learning rules, that take into account causal
information (i.e. which neuron triggers the firing of which); a
well-documented example is ``synaptic timing dependent plasticity''
(STDP) \cite{abbott2000} by which synaptic weights are either
reinforced or weakened depending on the relative spike timing between
the pre- and post-synaptic neuron.  STDP has been found to stabilize
the dynamics of neural networks and to maintain reproducible patterns
of causal neuronal avalanches \cite{Plenz2015}.  Thus, patterns of
activity, generated by neutral dynamics, and consisting on
mostly-non-interacting scale-free avalanches could be stored and
stabilized or ``ingrained'' by such a mechanism, allowing the network
to spontaneously generate a large set of attractors and a broad
dynamical repertoire, in a similar way in which cellular diversity
--stemming from underlying neutral dynamic of stem cells-- entails
functional advantages in epithelial tissues \cite{Simons}.  These
speculative ideas need to be much more carefully scrutinized, and we
plan to do so in forthcoming work.

Summing up, some of the existing confusion surrounding different types
of scale-invariance in neural activity can be rationalized in the
framework of neutral theories, posing new and fascinating questions
that may contribute to clarify the criticality hypothesis in the
cortex and its implications for function and learning.

\newpage
\section*{Materials and Methods}
{\footnotesize
\subsection*{Model for neural dynamics}
\label{appendix:millman}
The model of Millman \textit{et al.} \cite{Millman} consists of a
population of $N$ leaky integrate-and-fire excitatory neurons which
are randomly connected in a directed graph to, on average, other
{\footnotesize$K$} neurons in the population (i.e. forming a
Erd\H{o}s-R{\'e}nyi network \cite{Newman-review}).  External inputs,
{\footnotesize$I_{\text{e}}^k(t)$}, are Poisson-distributed with rate
{\footnotesize$f_e$} and internal inputs,
{\footnotesize$I_{\text{in}}^k(t)$}, are generated from spiking
neurons in the network ({\footnotesize$k$} accounts for the input
number).  Both internal and external currents are modeled by
exponentials functions of amplitude {\footnotesize$w_\text{e/in}$} and
characteristic time {\footnotesize$\tau_s$}, {\footnotesize
  $I_{\text{e/in}_i}^k(t)=w_\text{e/in}\exp(-(t-t_{s_i}^k)/\tau_s)$},
where {\footnotesize$t_{s_i}^k$} represents the corresponding spiking
time of neuron $i$.  Each individual neuron {\footnotesize$i$} is
described by a dynamical variable {\footnotesize$V_i$} representing
its membrane potential.  When this value reaches a threshold value
{\footnotesize$\theta$}, the neuron spikes and it may open --with
probability {\footnotesize$p_r$}-- each of its {\footnotesize$n_r$}
associated release sites per synapse, inducing a postsynaptic
current. After spiking, the membrane potential is reset to the resting
potential value, {\footnotesize$V_r$}, for a refractory period
{\footnotesize$\tau_{rp}$}, during which its dynamics is switched-off.
Synaptic depression is implemented by means of a dynamical ``utility''
variable {\footnotesize$U_{ij}(t)\in [0,1]$} (for neuron
{\footnotesize$i$} and release site {\footnotesize$j$}), which
modulates the release probability {\footnotesize$p_r\rightarrow
  U_{ij}p_r$}.  The membrane potential obeys the following equation:
{\footnotesize
\begin{equation}
  \dot{V}_i=-\frac{V_i-V_r}{RC}+\sum_{k}\frac{I^k_{\text{e}_i}(t)}{C}
  + \frac{1}{C}\sum_{\substack{i'\in n.n.(i)\\j,k}}\Theta(p_rU_{ i'j }
  (t_{s_{i'}}^k)-\zeta_{i'j}^k)
{I_{\text{in}_{i'}}^k(t)},\label{eq:Millman1}
\end{equation}
}
where {\footnotesize$R$} is the membrane resistance {\footnotesize$C$} its capacitance, 
{\footnotesize$k$} is the spike number, {\footnotesize$i'$} runs over presynaptic neurons linking to
{\footnotesize$i$}, and {\footnotesize$j'$} over its release sites; {\footnotesize$\zeta_{i'j'}^k$} is a uniform
random number in {\footnotesize$[0,1]$} and {\footnotesize$\Theta(x)$} the Heaviside step function.
On the other hand, the synaptic utility {\footnotesize$U_{ij}$} is set to {\footnotesize$0$}
immediately after a release and recovers exponentially to {\footnotesize$1$} at
constant rate, {\footnotesize$\tau_R$}:
{\footnotesize
\begin{equation}
\dot{U}_{ij}=\frac{1-U_{ij}}{\tau_R} -
\sum_{k}U_{ij}\Theta(p_r-\zeta_{ij}^k)\delta(t-t_{s_{i}}^k).
\label{eq:Millman2}
\end{equation}}
As equations \eqref{eq:Millman1} and \eqref{eq:Millman2} are linear
during successive events, they can be integrated exactly, which
allowed us to implement both synchronous (or clock-driven) and
asynchronous (or event driven) methods \cite{brette2007}, leading 
to essentially indistinguishable results.  When not specified,
model parameters were taken as in \cite{Millman}: {\footnotesize$K=7.5$, $n_r=6$, $R=2/3$ G$\Omega$,
  $C=30$ pF, $V_r=-70$ mV, $\theta=-50$ mV, $w_{\mathrm{e}}=95$ pA, $w_{\mathrm{in}}=50$
  pA, $p_r=0.25$, $\tau_{rp}=1$ ms, $\tau_s=5$ ms and $\tau_R=0.1$ s.}
We also studied versions of the model including inhibitory
couplings, but this did not alter the main conclusions (see SI Appendix S3).

\subsection*{Steady state density in the minimal model for activity propagation}
Neglecting fluctuations from finite size effects, the dynamics of the
total density of activity for the process described by
Eq. \eqref{eq:reaction} becomes deterministic in the limit
{\footnotesize$N\rightarrow\infty$}: {\footnotesize $ \dot \rho =
  (\lambda (1-\rho) - \mu) \rho + \epsilon(1-\rho)$}.  Its stationary
solution, {\footnotesize$\dot\rho=0$}, is {\footnotesize $\rho^* =
  (\lambda-\mu-\epsilon -
  \sqrt{4\epsilon\lambda+(\lambda-\mu-\epsilon)^2})/2\lambda$}, that,
up to first order in {\footnotesize$\epsilon$}, can be written as
{\footnotesize $\rho^*\simeq 1-\mu/\lambda+\epsilon \mu /( \lambda
  (\lambda-\mu) )$} in the active phase, {\footnotesize$\lambda>\mu$},
and {\footnotesize$\rho^*\simeq\epsilon/(\mu-\lambda)$} in the
quiescent one, {\footnotesize$\lambda<\mu$}.
}

\section*{Acknowledgments}
  We are grateful to the Spanish-MINECO for financial support (under
  grant FIS2013-43201-P; FEDER funds) and to P. Villegas, P. Moretti,
  S. Suweis and S. Vassanelli for extremely useful discussions and
  comments.

\newpage

\setlength{\voffset}{0cm}
\setlength{\hoffset}{0cm}


\section*{Supplementary material}

# Supporting Information: Neutral theory and scale-free neural dynamics

M. Martinello, J. Hidalgo, S. di Santo, A. Maritan, D. Plenz and M. A. Muñoz*

*E-mail: mamunoz@onsager.ugr.es

## SI Appendix S1: Brief summary of neutral theory

Consider a fully connected network with $N$ nodes (extensions to regular lattices, or more complex networks architectures are also possible, but we stick here to the simplest case) and a number of possible states (be these species, alleles, opinions, etc.). Each node adopts one of the possible states at every time. The dynamics proceeds as follows: at each time step, one randomly chosen individual is "invaded" by a copy of another neighboring node at uniform rate, i.e. common to all the individuals in the population independently of their species labels. Without loss of generality, we focus in the dynamics of a particular species, that we call $A$, and consider that any other individuals corresponds to species $B$, i.e. for simplicity we consider the case with just two species. This is nothing but the "voter model" (VM) [1, 2, 3, 4], also known as Moran process in the context of population dynamics and population genetics (see e.g. [5]). The VM has been profusely studied in the mathematical literature; some of its main relevant features are [1, 2, 3, 4, 5]: (i) it has no free parameters, (ii) it lacks any characteristic (length or time) scale and its dynamics exhibits scale-invariance, and (iii) it is characterized by purely noise-driven diffusive dynamics (see [3, 4] for more mathematical in-depth presentations).

Now, we derive the coarse-grained mean-field description of a voter model; similar derivations can be found in the literature [5]. For the sake of illustration, let us consider also a general model in which $\lambda_A$ (resp. $\lambda_B$) is the probability for $A$ (resp. $B$) to invade a site in state $B$ (resp. $A$), with $\lambda_A \neq \lambda_B$ in general; the VM dynamics is recovered imposing the neutral condition $\lambda_A = \lambda_B$.

As the system is saturated, the number of individuals for the other species is $n_B = N - n_A$ and the state of the system can be determined by the total number of individuals of $A$, $n_A$. The model can be expressed as a branching process [1], with transition rates $W(n_A \to n_A + 1) = \lambda_A n_B n_A / N$ and $W(n_A \to n_A - 1) = \lambda_B n_A n_B / N$. Using these rates, writing down the master equation for the probability of finding the system in a state $n_A$ at time $t$ –or alternatively with a density of individuals $A$, $\rho_A = n_A/N$– and performing a standard large $N$ expansion, one readily obtains the following Fokker-Planck equation:

$$\frac{\partial P(\rho_A, t)}{\partial t} = -(\lambda_A - \lambda_B)\frac{\partial}{\partial \rho_A}[\rho_A(1-\rho_A)P(\rho_A, t)] + \frac{\lambda_A + \lambda_B}{2N}\frac{\partial^2}{\partial \rho_A^2}[\rho_A(1-\rho_A)P(\rho_A, t)], \quad \textbf{(S1)}$$

or its equivalent (Itö) Langevin equation

$$\dot\rho_A = (\lambda_A - \lambda_B)\rho_A(1-\rho_A) + \sqrt{\frac{\lambda_A + \lambda_B}{N}\rho_A(1-\rho_A)}\eta(t), \quad \textbf{(S2)}$$

where $\eta$ is a zero-mean Gaussian white noise with $\langle \eta(t)\eta(t') \rangle = \delta(t-t')$.

The neutrality condition $\lambda_A = \lambda_B$ implies that the deterministic drift in Eq. (**S2**) vanishes thus $\langle \dot\rho_A \rangle = 0$, i.e. the average density of each species remains constant on average; their populations do not grow nor shrink on average, but they experience stochastic demographic changes as described by

$$\dot\rho_A = \sqrt{\rho_A(1-\rho_A)}\eta(\tilde{t}) \quad \textbf{(S3)}$$

where a factor $2\lambda_A/N$ has been absorbed into the new timescale $\tilde{t}$.

Observe that this last equation describes a stochastic process (random walk) with two absorbing barriers at 0 and 1. By neglecting the quadratic term in the noise (which is a valid approximation as far as the avalanche is small with respect to the much-larger system size), the avalanche-time exponent $\alpha = 2$ can be

deduced from the first-passage time (return to the origin) statistics of this random-walk process, and using simple scaling arguments one can also easily derive $\tau = 3/2$ for the avalanche-size distribution [6], i.e. one recovers the mean-field exponents of the voter model (neutral theory) class (see e.g. [7]). Needless to say that these same exponent values can also be computed by employing the more standard generating function formalism for a critical branching process (see, e.g. [1, 8]).

A difference between the VM dynamics and the contact-process-like one described in the main text is that in the VM the system is "saturated", in the sense that each single site is in one of the two possible opinions/alleles/species/labels/states, whereas in the model we study, some sites can be inactive, not belonging to any avalanche. However, looking at the dynamics of individual avalanches, such a difference becomes irrelevant.

# SI Appendix S2: Avalanche statistics for Down states in the model of Millman et al.

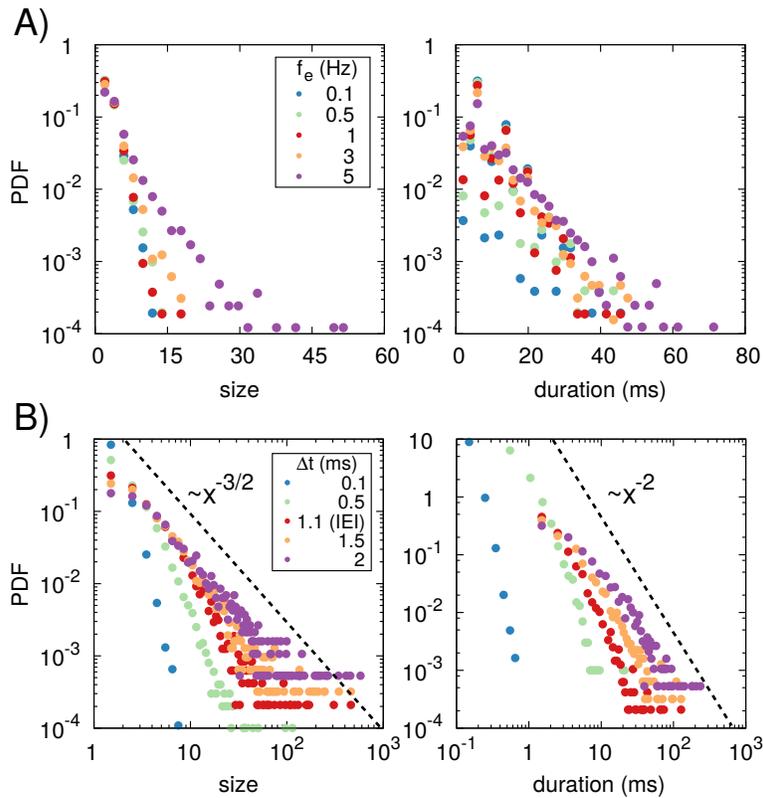

**Supplementary Figure S1**: Avalanche size and duration distributions relative to periods of low activity (Down states) for timeseries generated with the model of Millman *et al.* [9] using two different methods. Panel A (linear-logarithmic plot): "Causal" avalanches were defined using the same criterion as in [9], for several values of the external input $f_e$, confirming the observation that sizes and durations are exponentially distributed. Panel B (double logarithmic scale): "Time-correlated" avalanches, defined with the standard temporal binning method [10] (which ignores causal information), using five different time intervals $\Delta t$ to bin the data, including one coinciding with the average interevent interval (IEI) as usually done in the analyses of empirical data [10], for $f_e = 5$ Hz. In all cases, simulations were performed in a network of $N = 3000$ neurons (model parameters as in [9], see MM in main text).

# SI Appendix S3: Causal avalanches in the model with inhibitory synapses

Results presented in the main text are robust under the introduction of inhibitory synapses. Following [9], we run simulations of the model for which 20% of the neurons are initialized as inhibitory (so their output current has amplitude $-w_{\text{inh}}$) and, to keep the network balanced, each single neuron receives $k_i = k = 10$ inputs, 2 of which are from inhibitory ones. The introducing of inhibitory currents increases the coefficient of variation of spiking times, leading to enhanced variability. Still, as above, there exists a wide region of the parameter space, within the Up state (active phase) where causal avalanches keep showing scale-invariant behavior (see Fig. **S2**). More specifically, this happens whenever the amplitude of the inhibitory current is not too strong (so as to allow for the Up state to exist) and for values of the inhibitory synaptic timescale up to four times larger than the excitatory one.

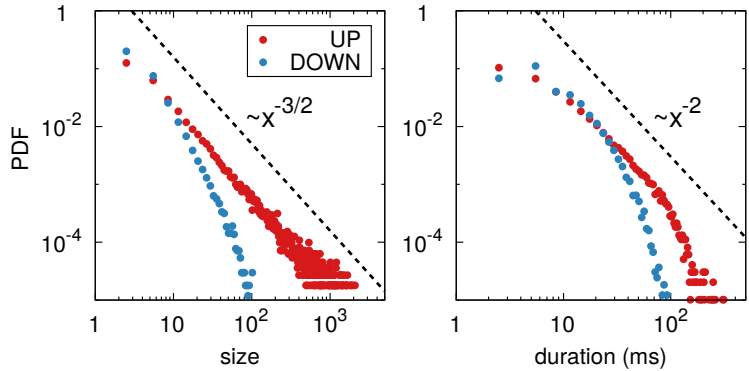

**Supplementary Figure S2**: "Causal" avalanche size (Left) and duration (Right) distributions in the model of Millman *et al.* including inhibitory synapses. The avalanche statistics exhibit a robust power-law scaling with the same exponents of a critical branching process (marked with dashed lines for comparison) with the Up state, while they are exponentials, with a characteristic scale in the Down state. Simulations were performed in a network of $N = 3000$ neurons setting the inhibition amplitude to $w_{\text{inh}} = 50$ pA, and other parameters as in [9] (see MM in main text).

# SI Appendix S4: Avalanche statistics in the simplified model

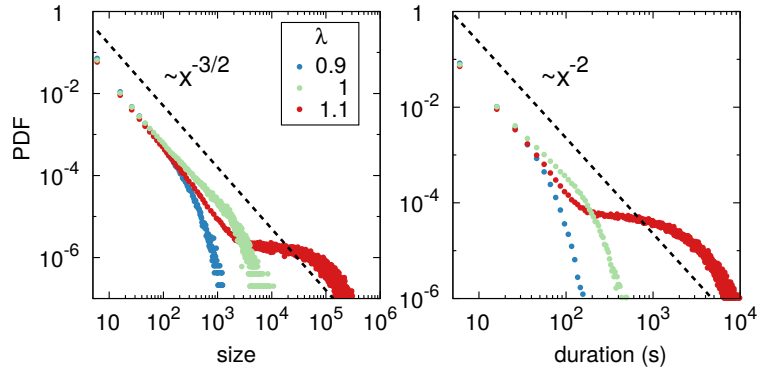

**Supplementary Figure S3**: Avalanche size (Right) and duration (Left) distributions in the simplified model for activity propagation when avalanches propagate in the network one at a time, i.e. without overlap between avalanches. This is done by setting $\epsilon = 0$ and introducing one single active node each time the activity stops. Distributions are plotted for different values of the activation rate $\lambda$ in the quiescent phase ($\lambda = 0.9$), active phase ($\lambda = 1.1$) and at criticality ($\lambda = 1$), illustrating the PDFs are not scale-invariant, except right at the critical point. Parameter values are: $\mu = 1$, $N = 10^3$.

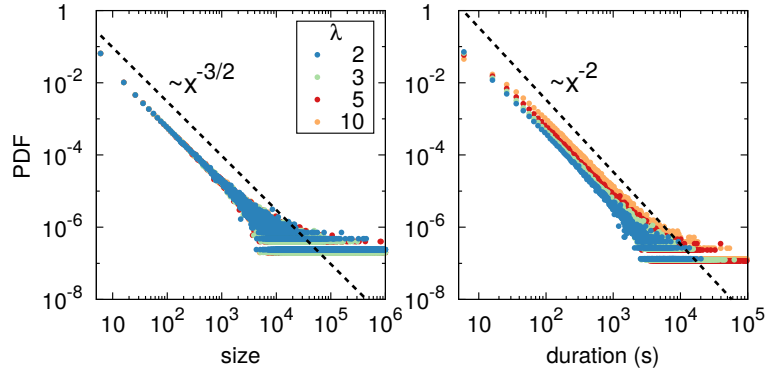

**Supplementary Figure S4**: "Causal" avalanche size (Right) and duration (Left) distributions in the simplified model for activity propagation when multiple avalanches coexist in the network, for different values of the activation rate $\lambda$ along the active (UP) phase (corresponding to stationary densities around $\rho^* = 0.5$, 0.66, 0.8 and 0.9, respectively). In all cases, avalanche statistics exhibit a robust power-law scaling with the same exponents of a critical branching process (marked with dashed lines for comparison). Increasing the spreading rate results in an enlargement of the mean duration of avalanches, with negligible effect on the size distributions. Parameter values are: $\mu = 1$, $\epsilon = 10^{-3}$, $N = 10^4$.



\end{document}